\documentclass{ostemp}
\usepackage{amsmath}
\usepackage{epsfig} % less modern

\usepackage{algorithm}
\usepackage{algorithmic}

\newcommand{\mypsfig}[2]{\psfig{silent=,figure=./{#1},width=#2\linewidth}} 
\newcommand{\ignore}[1]{}

\newcommand{\reals}{{I\!\!R}} % This is temporary, to allow texing on .166

\newcommand{\triplet}{\paren{p_i,p^+,p^-}}

\newcommand{\W}{{\bf{W}}}
\newcommand{\lW}{{l_{\W}}}

\newcommand{\Wi}{{\W_i}}
\newcommand{\WT}{{\W_0}}

\newcommand{\half}{\frac{_1}{^2}}
\newcommand{\dwiwj}{\|\Wi-\W_j\|_{Fro}^2}
\newcommand{\Ni}{\overline{{{{N}}}(i)}}
\newcommand{\Vold}{p(p^+-p^-)^T}
\newcommand{\Vp}{V}

\newcommand{\Vfro}{\|V\|^2}
\newcommand{\olddelta}{(p^+-p^-)}
\newcommand{\deltap}{\delta_p}
\newcommand{\avgwj}{\overline{\W_j}}
\newcommand{\dwiwjd}{\|\paren{\avgwj + {\Ne} \tau \Vp}-\W_j \|^2}

\newcommand{\oneminusbeta}{\paren{1-\beta}}

\newcommand{\weig}{\alpha_j}
\newcommand{\dwcwp}{\|\WT-\WT^{t-1}\|_{Fro}^2}
\newcommand{\dwcwpd}{\|\paren{\WT^{t-1} + \tau \Vp}-\WT^{t-1}\|^2}

\newcommand{\loss}{p^T \paren{\beta \Wi +\paren{1-\beta} \WT} \deltap}

\newcommand{\tauVhelper}{\paren{1 -\beta+\Ne\beta}}

\newcommand{\lossdowub}{\beta p^T\avgwj \deltap-\oneminusbeta p^T \WT^{t-1} \deltap }
\newcommand{\lossdowue}{- \tauVhelper \tau \Vfro}

\newcommand{\Ne}{N_e}

\newcommand{\newtauII}{\frac{l_{\Wi \WT} \paren{p_i,p^+,p^-}}{\tauVhelper\|\Vp\|^2 }}

\newcommand{\paren}[1]{\left({#1}\right)}
\newcommand{\bracket}[1]{\left[{#1}\right]}
\newcommand{\braces}[1]{\{{#1}\}}
\newcommand{\norm}[1]{\|{#1}\|}

\copyrightyear{2005}
\pubyear{2005}

\begin{document}
\firstpage{1}

\title{COLoR - Coordinated On-Line Rankers for Network Reconstruction}

\author[Sample \textit{et~al}]{Ossnat Bar-Shira\,$^{1}$ and Gal Chechik\,$^{1}$ }
\address{$^{1}$The Gonda Brain Research Center, Bar Ilan University 52900 Ramat-Gan, Israel.\\
}
\maketitle

\maketitle

\begin{abstract}

\section{Motivation:}
Predicting protein interactions is one of the more interesting challenges of the post-genomic era.
Many algorithms address this problem as a binary classification problem: 
given two proteins represented as two vectors of features, predict if they interact or not.
Importantly however, computational predictions are only one
component of a larger framework for identifying PPI. The most
promising candidate pairs can be validated experimentally by testing
if they physical bind to each other.
Since these experiments are
more costly and error prone, the computational predictions serve as
a filter, aimed to produce a small number of highly promising candidates.
Here we propose to address this problem as a ranking problem:
given a network with known interactions, rank all unknown pairs based
on the likelihood of their interactions. 

In this paper we propose a ranking algorithm that
trains multiple inter-connected models using a passive aggressive on-line approach.
We show good results predicting protein-protein interactions for post synaptic density  PPI network. 
We compare the precision of the ranking algorithm with local classifiers \citep{Bleakley07} and classic SVM \citep{Vapnik98}
Though the ranking algorithm outperforms the classic SVM classification, its performance is inferior compared to the local supervised method.

\section{Availability:}
Interaction inference package is available upon request from the authors. 
\section{Contact:} \href{asnat.bar-shira@live.biu.ac.il}{asnat.bar-shira@live.biu.ac.il }, \href{ gal.chechik@biu.ac.il}{gal.chechik@biu.ac.il}
\end{abstract}

\section{Introduction}
Revealing the interactions between groups of proteins will considerably contribute to our understanding of intracellular processes such as signal transduction, trafficking, or regularization.
The fastest way to explore interactions between proteins is via high throughput experiments.
In the last decade many high-throughput methods, such as yeast two-hybrid \citep{Fields89}, which detects protein interactions or mass spectrometry \citep{Ho02}, that identifies components of protein complexes, were developed.
These methods systematically probe interactions in a large group of proteins. But the data obtained by such studies is partial, and susceptible to under- and over- detection \citep{Qi06}.
Moreover, different methods yield different interactomes. In yeast for example, Von Mering {\it et~al}. have found that only 2400 out of 80,000 interactions detected by large-scale approaches were supported by more than one method.
A biological solution for this ambiguousness is to use small scale methods to check the proteins one by one, validate the detected connections and find out the undetected ones. This is of course a tedious, time consuming job, which will take years if extensively done.
A more feasible way is to introduce an intermediate step, in which interactions will be inferred {\it in silico}. 
In this setting, computational methods put forward a short list of most probable binds, which can then be experimentally verified.

The wealth of biological data beside the interactomes themselves, hearten the use of supervised learning methods. Data sets such as gene expression \citep{Eisen98} , protein localization \citep{Huh03}, signatures \citep{Apweiler01} or phylogenetic profiles \citep{Pellegrini99}, can readily serve as bags of features for the inspected proteins.

The most common approach to edge prediction via supervised learning is to train a binary classifier from available datasets. And use it to infer unknown interactions within this set of proteins \citep{Yamanishi04},\citep{Ben-Hur05}. However, the features that are predictive, may differ across different families of proteins, or even
change dramatically from one protein pair to the other. For example,
gene-expression features measured in under amino-acid starvation
condition may be very predictive for bio-synthesis proteins, but not
for mitochondrial proteins. As a result, learning a single unified
model across the full network may over generalize. 

A possible solution is to train a separate classifier for each protein \citep{Bleakley07}, which considerably narrows the amount of data that can be used to train each classifier.
A possible alternative is to train a set of dependent classifiers that share data between associated proteins.
For this purpose we developed COLor, a coordinated local ranker.
COLor treats the edge prediction as a multi task 
ranking problem. It defines a separate learning problem for each protein in the
network. For regularization, it constrains models of neighboring proteins
to be similar, yielding a set of models that smoothly varies across
the network. 
Since the number of models is large (equal to the number of nodes), we
take an on-line large margin approach that is based on the
Passive-Aggressive family of models, and scales to handle thousands of models.
In this paper we examine ranking vs. global and local classification algorithms. We found COLoR to be much more precise than the global SVM. Unfortunately, for mid-size PPI networks, which are our networks of interest, the local classifiers outperform COLoR.

\begin{methods}
\section{Methods}

\subsection{The learning problem}

Let $P$ be a set of proteins, $p_1,\ldots,p_n \in \reals^d$, and  $G=
\paren{P,E}$ be a graph that defines their pairwise interactions.
Each edge in the graph $e(p_i,p_j)\in\{0,1\}$ is a binary variable
with a value $1$ iff $p_i$ and $p_j$ interact.

Our goal is to learn a scoring function $S_w(p_i,p_j)$ with parameter w, that assigns a
higher score to pairs that interact.
\begin{eqnarray}
  &  & S_{\W}(p,p^+) \ge S_{\W}(p,p^-), 
\\ \nonumber
& &\quad \forall \triplet ,
  e(p,p^+)\in E, e(p,p^-)\notin E, \forall{p_i}\in{V}
\end{eqnarray}
In what comes below, we focus on bi-linear scoring functions of the
form $ S_{\W}(p,q) = p^T \W q$.

Typically, a single scoring function is trained that is common to all
pairs. However, the way features predict if two proteins interact may
vary quite substantially in different parts of the network. 

To capture such variability we learn multiple models (scoring
functions): a specific model for each node in the network,
and a global scoring function that is common to all vertices. The 
combined scoring function will be -
\begin{equation}
   S_{\W}(p,q) = \beta p^T \Wi q + (1-\beta) p^T \WT q
\end{equation}
Where $\WT$ is the global learner that is common to
all interactions, $\Wi$ is a local learner specific to the $i^{th}$
node, and $\beta$ is a trade-off parameter, that weights the
contribution of the local and global models. When $\beta = 0$, the
model reduces to the single-task learning problem. Following  \citep{Grangier08} and \citep{Chechik10}, 
we minimize the following hinge loss for every triplet
$\triplet$
\begin{equation}
  \lW \triplet  =  \max \paren{0, 1 - S_{\W}(p,p^+) + S_{\W}(p,p^-)}
\end{equation}

Naturally, the local models must be regularized so learning can
generalize across pairs. We achieve this by defining a set of 
$\alpha_{ij}$ forcing the two models $\Wi$ and $\W_j$ to be close in $L_2$
terms. This defines penalty terms of the form
$\alpha_{ij}\norm{\Wi-\W_j}$. When prior knowledge is available about
which nodes should be similar, it can be used to set the
$\alpha$'s. Alternatively, we set the $\alpha$'s based on the known network
structure $\alpha_{ij} = e(p_i,p_j)$. This way, connected nodes
are pushed to have similar models.

The multi-task problem of optimizing all $\Wi$ jointly is a large
problem, since typical PPI networks include hundreds of nodes and edges. We
therefore take an on-line approach for optimizing all $\Wi$. In this
setting, it is natural to add another regularization term that forces
$\W$ at each step to remain close to its previous value $\norm{\Wi^t
- \Wi^{t-1}}$. Together, we therefore define an "influence set" for
each vertex model $\Wi$ that contains the neighbors of $p_i$ and an
additional pseudo-neighbor $\Wi^{t-1}$ which holds $i$'s history.  We
denote by $\Ni$ the extended set of "neighbors'', $\Ni = {\braces{j |e_{ij}=1}} \cup i_{history}$ where $i_{history}$ 
refers to same node in the previous round. We set the weight of the pseudo-neighbor to $\alpha=1$.

\subsection{A ranking algorithm for local models}
% --------------------------------------------
We describe an on-line multi-task learning algorithm based on the family
of passive aggressive algorithms introduced by \citep{Crammer06}. First,
all $\Wi$'s are initialized at $\Wi = I$. Then, at each iteration $t$,
a protein $p_i$ is sampled, together with a protein $p^+$ s.t. $e(p,p^+) \in E$ and a protein $p^-$
that is not connected to $p$. This provides a triple $\triplet$ for which
we define the following optimization problem
\begin{eqnarray}
  \label{eqpallm}
     \min_{\W} &&
	\frac{_\beta}{^{2\Ne}}\sum_{j \in \Ni}^n \weig \dwiwj 
	\\ \nonumber
	& &  
	+ \frac{_{1-\beta}}{^2} \dwcwp + C \xi_i  \\ \nonumber
	{\rm s.t.} & & \lW \triplet \leq \xi_i \\ \nonumber
	& & \xi_i \ge 0
\end{eqnarray}

We follow \citep{Crammer06} to develop an algorithm for solving
\eqref{eqpallm}.  If $l_{\W^t}=0$ no update is needed. Otherwise, we
define the Lagrangian
\begin{eqnarray}
&\mathcal{L}&\paren{W_i,W_T,\tau,\xi,\lambda} = 	
	\\ \nonumber 
        && \frac{_\beta}{^{2\Ne}} \sum_{j\in \Ni}^n \weig \dwiwj   \\ \nonumber
        && + \frac{_{1-\beta}}{^2} \dwcwp + C \xi_i  \\ \nonumber
        && +\tau \bracket{1-\xi_i-\loss}-\lambda\xi_i 
	\\ \nonumber
\end{eqnarray}
where $\delta_p = \olddelta$, $\tau \ge 0$ and $\lambda \ge 0$ are
Lagrange multipliers. To find the optimal solution, we equate
$\frac{\partial\mathcal{L}}{\partial \Wi}$ and
$\frac{\partial\mathcal{L}}{\partial \WT}$ to zero, and obtain 
\begin{eqnarray}
   % \frac{\partial\mathcal{L}}{\partial \Wi} & = & 
    % \frac{1}{\Ne} \beta \paren{\Wi -\avgwj} -\beta \tau \Vp \; = \; 0
    %     \\  \nonumber
    \Wi & = & \avgwj + \Ne \tau \Vp \\ \nonumber
    % \frac{\partial\mathcal{L}}{\partial \WT} 
    % & = & \oneminusbeta \WT - \oneminusbeta \WT^{t-1} + \tau \Vp \; = \; 0 \\  \nonumber
    \WT & = & \WT^{t-1} + \tau \Vp
\end{eqnarray}
where $\avgwj$ is the average of the neighbors W's, 
$\Ne$ is the number of neighbors, and
$\Vp = \frac{\partial{p^T \W \deltap}}{\partial \W} =\Vold$.

Deriving the Lagrangian with respect to $\xi$ and setting it to zero yields
\begin{eqnarray}
	\frac{\partial\mathcal{L}}{\partial \xi_i} &=& C-\tau-\lambda = 0 \\ \nonumber
\end{eqnarray}
Plugging the above back into (5), and taking the derivative w.r.t. $\tau$

\begin{eqnarray}
\mathcal{L}\paren{\tau}& = & \frac{1}{\Ne} \beta \sum_{j\in \Ni}^n \weig \half \dwiwjd  \\ \nonumber
	& &  + \oneminusbeta \half \dwcwpd + C \xi_i \\ \nonumber
%	& &  +\tau \lossdowut \\ \nonumber
         & & +\tau\lossdowub \\ \nonumber
         & & \lossdowue\\ \nonumber
\end{eqnarray}
which gives us the optimal $\tau$

\begin{eqnarray}
	\tau & = & min\braces{C, \newtauII}
\end{eqnarray}

We name this algorithm COLoR, Coordinated On-line Local Rankers. Algorithm1 presents the pseudo code. \\

\begin{algorithm}\label{Table 1}
 \caption{COLoR - Online algorithm for learning coordinated models}
 {\bf Initialization:} Initialize $W_T^0,Wi^0 = 1$.\\
  {\bf Iterations:}
 \\
 \hspace{10pt}
\begin{tabular}{l}
 \hspace{-5pt}{\bf  repeat: } \\
 \hspace{12pt}Sample three proteins such that - \\
	 \hspace{12pt}$Pr(e(p_i,p^+)=1) \ge Pr(e(p_i,p^+)=1)$ \\
	\hspace{12pt}Update $\Wi  =  \avgwj + \Ne \tau_i \Vp$, $\WT^i = \WT^{t-1} + \tau_i \Vp $ \\
	\hspace{12pt}Where $\tau_i  =  min\braces{C, \newtauII}$\\
	\hspace{12pt}and $\Vp  =  \frac{\partial{p^T \W \deltap}}{\partial \W} \; = \; \Vold $
\\

\end{tabular}
\end{algorithm}

\section{Experiments}
% =============================

We tested the algorithm on a network of PPIs describing interactions in the post synaptic density.  The network was constructed from trusted scientific reports that describe interactions between proteins in the post synaptic density (PSD).
Proteins that inherently lacked data for one or more feature sets, For example, proteins which genes are not included on the mouse expression genechip, were removed. The resulting network included 211 interactions between 114 proteins. To integrate the various sources, and to be compatible with other data sources (NCBI geo gene expression for example), networks vertices were represented by their gene symbols as specified in the Mouse Nomenclature guidelines (http://www.informatics.jax.org/mgihome/nomen/gene.shtml).

\subsection{Protein features}
% =============================
\subsubsection{Gene Expression}
We downloaded 459 microarray expression profiles from NCBI Geo (http://www.ncbi.nlm.nih.gov/geo/), all belong to NCBI GPL81 platform (Mus Musculus Affymetrix Murine Genome U74 Version) which measures expression profiles of 12488 genes. We retrieved datasets that hold the results of brain related experiments only. The expression data for each experiment results column was normalized using Cox-Box transformation and scaled using zero mean and unit variance. Data was downloaded from NCBI geo at Mar 28 2010.

\subsubsection{Phylogenic data}
Pairwise ortholog maps of 99 species were downloaded from the Inparanoid database (http://inparanoid.cgb.ki.se/)
For each gene we calculated ortho-score by multiplying of the gene's confidence score and the confidence level of this paralog cluster (ortholog group bootstrap value). We created a table of all Mus musculus genes, as given in MGI (http://www.informatics.jax.org/), and their ortho-scores against all other 98 orthologs. The orthoXML files were downloaded on December1st 2010.

\subsubsection{Protein domains and signature annotations}
We downloaded data from two databases, Interpro (http://www.ebi.ac.uk/interpro/), an integrative database of predictive models (signatures), and Pfam (http://pfam.sanger.ac.uk/), a repository of protein domains.
In Pfam, we used the high-quality manually curated Pfam-A domains only.  
Overall we used 122 Pfam domains and 21178 signatures. 
XML files were downloaded from Pfam on June 2010, and from Interpro on January 2011.
We used TF-IDF, a procedure borrowed from information retrieval \citep{Salton86}, to represent the domains and signatures by their weights. In our context, the domains or signatures serve as "terms", proteins as "documents" and the entire dataset as "corpus".

\subsubsection{Co-expression across brain structures}
We retrieved expression levels per structure, per gene, from the Alan Brain Atlas (http://www.brain-map.org/), which report expression levels across 17 different brain structures.  For each gene we built a vector of 17 entries, each represent expression levels in the different brain structures. Search is done on-line.

\subsection{algorithms comparison}
% ================================
We compared the performance of COLoR with global and local classifiers.
Both classifiers are based on support vector machines \citep{Vapnik98}. The global SVM approach trains a single prediction model for the whole network. Local SVMs \citep{Bleakley07} trains an independent model for each vertex.
To estimate the accuracy of the three approaches, we evaluate their predictions on held out data that was not used during training. Specifically, we use 5-fold cross-validation where at each fold, 80 percent of the proteins are used to train edge predictors and the remaining 20 percent are used to evaluate the precision of the learned classifier. Given a trained model, we used it to predict interactions between all candidate pairs of proteins and rank the pairs by the likelihood that they interact. We then computed the precision (fraction of truly interacting proteins) within the top-k ranked pairs.
Figure 1 depicts the precision at top-k as a function of k for all the approaches. The local SVM approach is the most precise, COLoR is not as good, but it is more precise than the Global SVM at the top 40 ranked predictions. 

\begin{figure} [h!]

  \centerline{ 
  \mypsfig{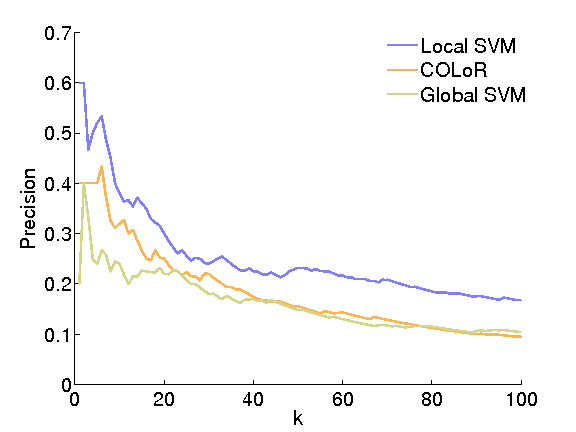}{1} 
  }
  \caption{\label{Figure:01} Precision at top k - COLoR, Global and Local SVMs.  }
\end{figure}

\subsection{features comparison}
% ==============================
In order to evaluate the predictive power of different features, we examined a collection of microarray results of brain related experiments from NCBI GEO \citep{Edgar02}, domains from Pfam \citep{Bateman04}, signitures from the Interpro \citep{Apweiler01}, ortholog maps from Inparanoid \citep{OʼBrien05}, and gene expression across brain structures from the Alan Brain Atlas \citep{Lein07}. 
We found that the best precision was achieved when combining expression, domains, signatures and phylogenetic data. (Figure 2). Data from the Alan brain Atlas was non-predictive by itself, and obstructed the classification when combined with the other data sets. 

\begin{figure} [h!]

   \centerline{ 
  \mypsfig{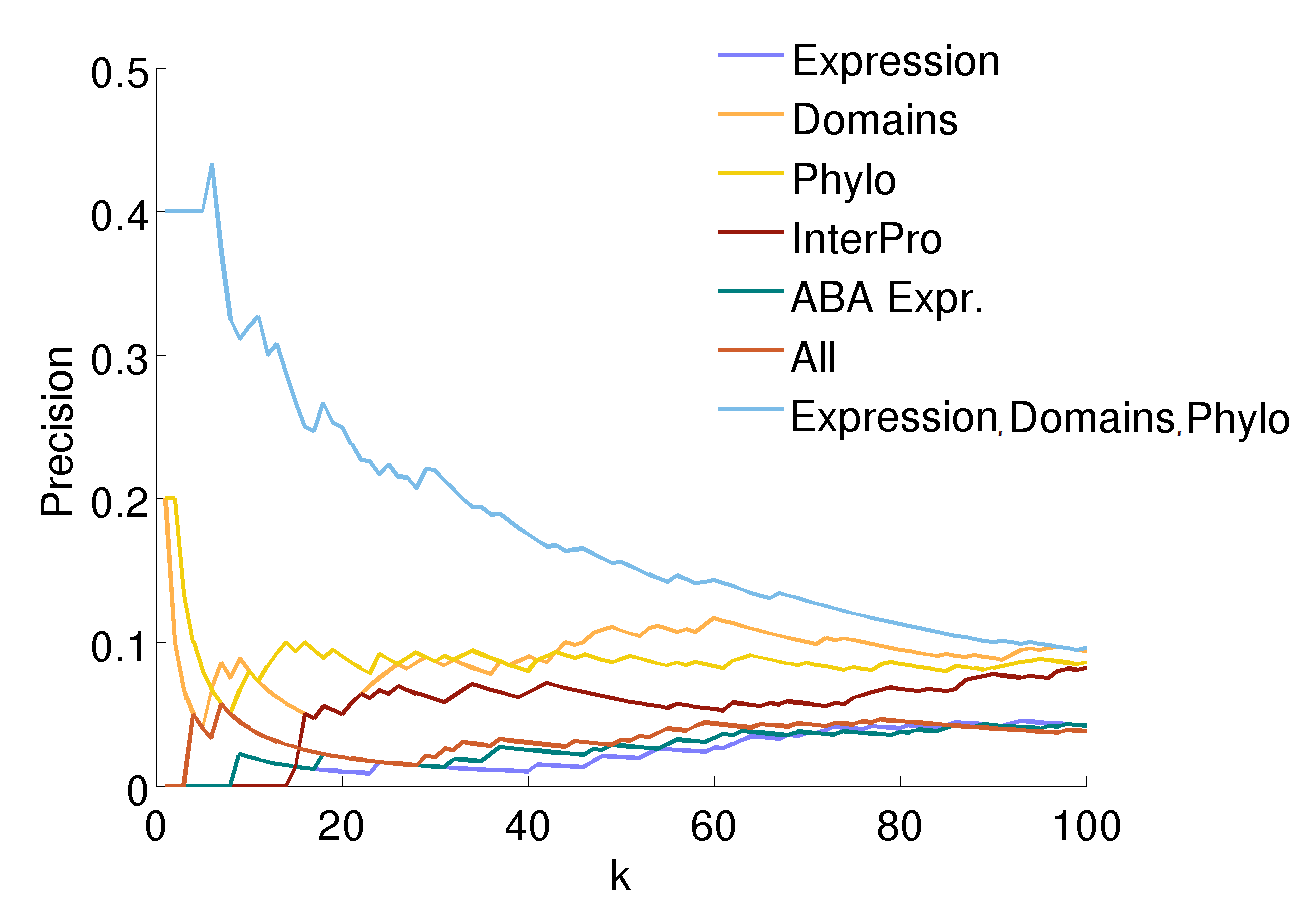}{1} 
  }
  \caption{\label{Figure:02} COLoR - Precision at top k for various features. }
\end{figure}

The same feature composition was proved to be the most predictive for the Global and Local SVMs as well (Figures 3 and 4 respectively)

\begin{figure} [h!]
  \centerline{ 
  \mypsfig{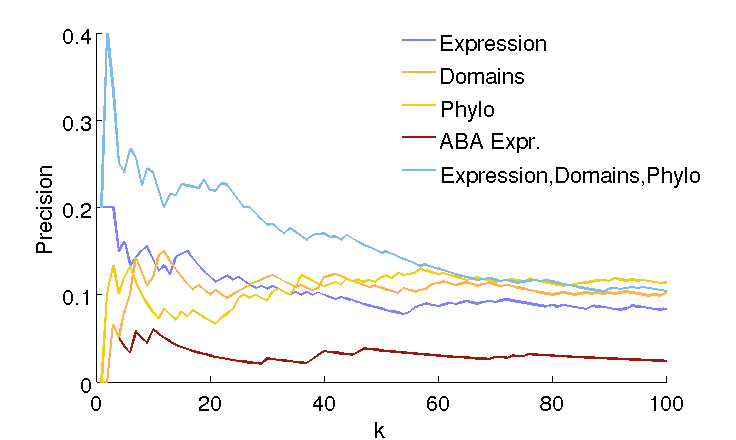}{1} 
  }
  \caption{\label{Figure:03} Global SVM - Precision at top k for various features.  }
\end{figure}

\begin{figure} [h!]
  \centerline{ 
  \mypsfig{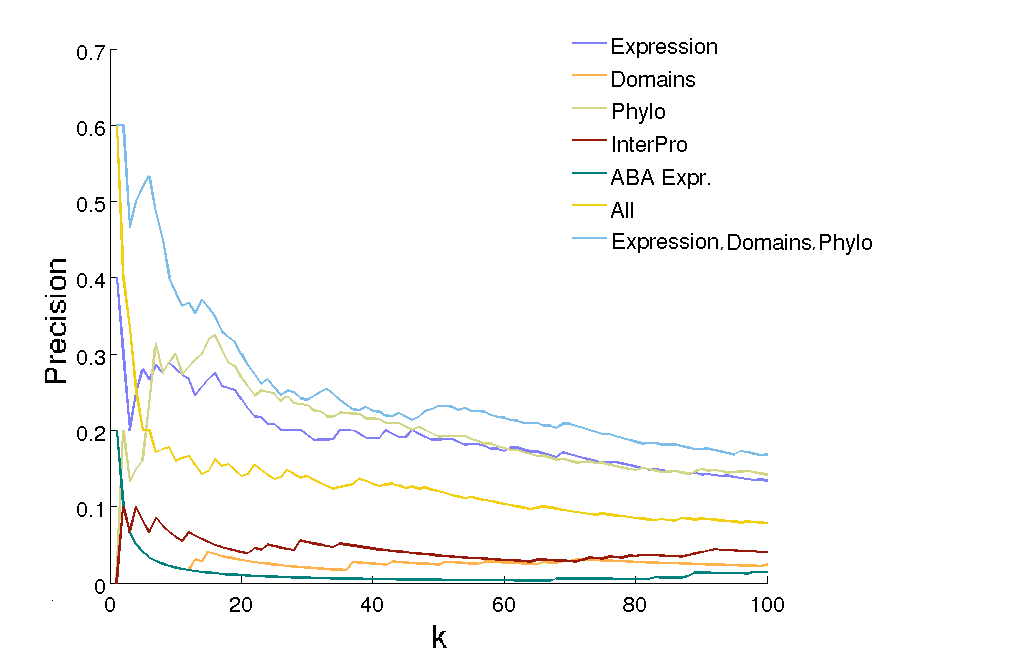}{1} 
  }
  \caption{\label{Figure:04} Local SVM - Precision at top k for various feature vectors.}
\end{figure}

\end{methods}

\end{document}